%% file: Wrapper_ATLAS.tex
\RequirePackage{latex/atlaslatexpath}

\documentclass[cernpreprint, texlive=2020, UKenglish, orcidlogo, cmslogo=true]{atlasdoc}

\DeclareOldFontCommand{\rm}{\normalfont\rmfamily}{\mathrm}
\DeclareOldFontCommand{\sf}{\normalfont\sffamily}{\mathsf}
\DeclareOldFontCommand{\tt}{\normalfont\ttfamily}{\mathtt}
\DeclareOldFontCommand{\bf}{\normalfont\bfseries}{\mathbf}
\DeclareOldFontCommand{\it}{\normalfont\itshape}{\mathit}
\DeclareOldFontCommand{\sl}{\normalfont\slshape}{\@nomath\sl}
\DeclareOldFontCommand{\sc}{\normalfont\scshape}{\@nomath\sc}

\usepackage{atlasphysics}
\usepackage{atlasjournal}

\usepackage{mathrsfs}
\usepackage{xspace}
\usepackage{comment}
\usepackage{booktabs} 
\usepackage{graphicx}
\usepackage{multirow}
\usepackage{hyperref}
\usepackage{rotating}
\usepackage{url}
\usepackage{verbatim}
\usepackage{amsmath}
\usepackage{bm}
\usepackage{color}
\usepackage{multicol}
\usepackage{caption}
\usepackage{subcaption}
\usepackage{placeins}

\graphicspath{{figures/}{logos/}}
\input{packages_and_commands.tex}

\AtlasTitle{Combination of inclusive top-quark pair production cross-section measurements using ATLAS and CMS data at $\sqrt{s}= 7$~and 8~\TeV}
\PreprintIdNumber{CERN-EP-2021-222}

\AtlasJournal{\JHEP}
\AtlasJournalRef{JHEP 07 (2023) 213}
\AtlasDOI{10.1007/JHEP07(2023)213}

\AtlasAbstract{ %
A combination of measurements of the inclusive top-quark pair production cross-section performed by ATLAS and CMS  in proton--proton collisions at centre-of-mass energies of 7~and 8~\TeV\ at the LHC is presented. The cross-sections are obtained using top-quark pair decays with an opposite-charge electron--muon pair in the final state and with data corresponding to an integrated luminosity of about 5\fbinv at $\sqrt{s}=7~\TeV$ and about 20\fbinv at $\sqrt{s}=8~\TeV$ for each experiment. The combined cross-sections are determined to be $\ttxvalcombs \pm \ttxtotpcombs$~pb at $\sqrt{s}=7~\TeV$ and $\ttxvalcombe^{+\ttxtotpcombe}_{-\ttxtotmcombe}$~pb at $\sqrt{s}=8~\TeV$ with a correlation of  \rhocomb, using a reference top-quark mass value of 172.5~\GeV. 
The ratio of the combined cross-sections is determined to be $R_{8/7}= \ttxratio\pm \ttxratioerr$. The combined measured cross-sections and their ratio agree well with theory calculations using several parton distribution function (PDF) sets. The values of the top-quark pole mass (with the strong coupling fixed at 0.118) and the strong coupling (with the top-quark pole mass fixed at 172.5~\GeV) are extracted from the combined results by fitting a next-to-next-to-leading-order plus next-to-next-to-leading-log QCD prediction to the measurements. Using a version of the NNPDF3.1  PDF set containing no top-quark measurements, the results obtained are $\mtpole = \massNNascorrnotop^{+\massuncNNascorrnotopup}_{-\massuncNNascorrnotopdown}$~\GeV\ and $\alphas(m_Z)= \alphasNNascorrnotop^{+ \alphasuncNNascorrnotopup}_{-\alphasuncNNascorrnotopdown}$.}

\begin{document} 
\maketitle

\input{text.tex}
%
%
\section*{Acknowledgements}
\input{Acknowledgements_ATLAS}
\input{CMS-TOP-18-014-acks}

\clearpage
\bibliographystyle{latex/atlasnote}
\bibliography{bibtex/bib/Theory.bib,bibtex/bib/ATLAS.bib,bibtex/bib/CMS.bib,bibtex/bib/CMS_default.bib}

\clearpage
\input{atlas_authlist}
\clearpage
\input{TOP-18-014-public-authorlist.tex}

\end{document}

%% file: text.tex
\section{Introduction}
\label{sec:introduction}

The top quark is the most massive known fundamental particle, with a mass close to the scale of the electroweak symmetry breaking~\cite{PhysRevD.98.030001}. Studying the production and decay of the top quark in proton--proton ($pp$) collisions is a crucial element of the CERN LHC physics programme and provides precise tests of the Standard Model (SM). At the LHC, top quarks are produced mostly in quark--antiquark pairs (\ttbar). For this production mode, precise predictions at next-to-next-to-leading order (NNLO) in perturbative quantum chromodynamics (QCD) including resummation of next-to-next-to-leading-log (NNLL) soft-gluon terms are available~\cite{topxtheo1,topxtheo2,topxtheo3,topxtheo4,topxtheo5,Catani_2019}. Consequently, precise measurements of the {\ttbar} production cross-section (\xtt)   may reveal contributions from non-SM processes that modify \xtt, such as those in supersymmetric models with R-parity conservation~\cite{FARRAR1978575}. Moreover, measurements of  \xtt can provide constraints on essential parameters of the SM, such as the top-quark pole mass (\mtpole) and the strong coupling (\alphas), and on parton distribution functions (PDFs). 
\medskip

The predicted \ttbar\ production cross-section in $pp$ collisions depends on the PDF set. Table~\ref{t:predxsecs} shows the predictions for the NNLO PDF sets CT14~\cite{CT14}, MMHT14~\cite{MMHT2014}, and NNPDF3.1\textsc{nnlo}\_ascorr\_notop (referred to as NNPDF3.1\_a)~\cite{NNPDF31ascorrnotop}.\footnote{The PDF sets NNPDF3.1\_a and CT14 do not include any top-quark-related measurements, such that any potential bias can be avoided when using them together with the combined cross-sections to extract a top-quark mass and the strong coupling. The MMHT14 PDF set does include information from the \ttbar\ cross-section at the LHC (but no alternative PDF sets were provided by the authors).} The precision of the prediction depends on the uncertainties in the PDF set, the dependence of the PDF set and the calculation on \alphas , and the choice of factorisation and renormalisation scales. These predictions are calculated at NNLO+NNLL accuracy in QCD for a top-quark pole mass of 172.5~\GeV\ with \toppp~\cite{toppp}. 
The quoted scale uncertainty is derived from the envelope obtained by independently varying the renormalisation and factorisation scales from \mtpole by factors of $0.5$ and $2.0$. Variations where these scales differ from each other by a factor of more than two are not included.
\medskip

\begin{table}[htp]
\caption{
Predicted \ttbar\ production cross-sections at different centre-of-mass energies and for different PDF sets. The uncertainties comprise PDF and  \alphas\ uncertainties as well as uncertainties in the renormalisation and factorisation scales. The ratio of the predicted \ttbar\ production cross-sections at $\sqrt{s}=8~\TeV$ and 7~\TeV, $R_{8/7}$, is also shown. NNPDF3.1\_a is a version of this PDF set containing no top-quark measurements.
\label{t:predxsecs}
}
\vspace{0.2cm}
\centering
\renewcommand{\arraystretch}{1.1}
\begin{tabular}{lccc}
\hline
PDF set  &  \xtt  ($\sqrt{s}=7~\TeV$) [pb] & \xtt ($\sqrt{s}=8~\TeV$) [pb] &$R_{8/7}$ \\
\hline
CT14 &  181.7 $^{+10.6}_{-10.3}$   &  258.9  $^{+13.8}_{-14.3}$ & 1.425 $^{+0.007}_{-0.008}$ \\
MMHT14 & 181.2 $^{+9.6}_{-10.3}$  &  258.1 $^{+12.8}_{-14.1}$  & 1.424 $^{+0.005}_{-0.004}$ \\
NNPDF3.1\_a & 178.8 $^{+7.8}_{-8.8}$ &  255.3  $^{+10.6}_{-12.2}$  & 1.428 $^{+0.005}_{-0.004}$ \\
\hline
\end{tabular}
\end{table}

Measurements of \xtt were performed previously by the ATLAS~\cite{ATLASpaper}  and CMS~\cite{CMSpaper} Collaborations at $\sqrt{s}=7~\TeV$ and 8~\TeV\ in various decay channels~\cite{ATLASttbar1, ATLAS:2015zgd, ATLAS:2014ixi, ATLAS:2012qtn, CMSttbar, CMS:2013yjt, CMS:2013nie,  CMS:2012ahl, ATLAS:2017wvi,  ATLAS:2017jkf, CMS:2016csa, CMS:2014btv, CMS:2013hon, CMS:2015auz}.
Each top quark decays almost exclusively into a \W boson and a \bq-quark. The subsequent decays of the \W bosons define the \ttbar\ final-state topology to be either fully hadronic, with one, or with two leptonic decays.
In this paper, ATLAS and CMS present a combination of their most precise measurements of \xtt in $pp$ collisions at $\sqrt{s}=7$ and 8~\TeV, obtained using \ttbar\ decays into electron--muon ($e\mu$) pairs~\cite{ATLASttbar1, CMSttbar}. 
The final-state topology of this decay mode is defined by the two leptons of opposite charge and different flavour, two jets which are identified as originating from final-state $b$-quarks ($b$-tagged jets), and missing transverse momentum from the undetected neutrinos. This final state also includes small contributions from $\W\rightarrow\tau\nu_\tau$ with subsequent leptonic $\tau$ decays.
The combined \xtt values are then used to extract  \mtpole and \alphas\ through comparison with NNLO+NNLL predictions for various PDF sets.
\medskip

The individual input measurements are briefly summarised in Sections~\ref{sec:atlas_meas} and~\ref{sec:cms_meas}. The combination method and the correlation assumptions used for the systematic uncertainties from ATLAS and CMS are described in Section~\ref{sec:combination}. The combined results are presented in Section~\ref{sec:results_and_mass}, along with the extraction of \mtpole and \alphas. The summary and conclusions can be found in Section~\ref{sec:conclusions}.

\section{ATLAS measurements}
\label{sec:atlas_meas}
The most precise measurements of the inclusive \ttbar\ cross-section in $pp$ collisions at $\sqrt{s}=7$ and 8~\TeV\ from the ATLAS Collaboration used events with an opposite-charge $e\mu$ pair and one or two $b$-tagged jets~\cite{ATLASttbar1}. The \ttbar\ production cross-section was determined by counting the numbers of opposite-charge $e\mu$ events formed from an isolated electron and isolated muon with exactly one ($N_1$) or with exactly two ($N_2$) $b$-tagged jets. A working point of 70\% efficiency for tagging $b$-jets from top-quark decays, corresponding to a rejection factor of about 140 against light-quark and gluon jets, was used. 
The two event counts can be expressed as
\begin{eqnarray}
N_1 & = & L \xtt\ \epsem 2\epsb (1-\cb\epsb) + \nib \nonumber ,\\
N_2 & = & L \xtt\ \epsem \cb\epsb^2 + \niib ,\label{e:tags}
\end{eqnarray}
where $L$ is the integrated luminosity of the sample and \epsem\ is the
efficiency for a \ttbar\ event to pass the opposite-charge $e\mu$ preselection.
The combined probability for a jet from the quark $q$ in the $t\rightarrow Wq$ decay to fall within the acceptance of the detector, be reconstructed as a jet with transverse momentum $\pt>25$~\GeV\ and pseudorapidity $|\eta|<2.5$, and be tagged as a $b$-jet, is denoted by \epsb.  If the decays of the two top quarks and the subsequent reconstruction of the two $b$-tagged jets are completely independent, the probability to tag both $b$-jets is given by $\epsb^2$. In practice, small correlations are present for both physical and instrumental reasons, and these are taken into account with the $b$-tagging correlation factor \cb . This correlation factor is close to unity such that a value greater than one corresponds to a positive correlation (i.e., where a second jet is more likely to be selected if the first one is already selected).
The background sources also contribute to the event counts $N_1$ and $N_2$. These contributions are represented by the terms \nib\ and \niib. 
The preselection efficiency \epsem\ and tagging correlation \cb\ were taken from \ttbar\ event simulation, assuming a top-quark mass of 172.5~\GeV, and the background contributions \nib\ and \niib\ were estimated using a combination of simulation and data-driven methods, allowing the two equations in Eq.~(\ref{e:tags}) to be solved for \xtt\ and \epsb, independently for 7 and 8~\TeV. The effect of the small contributions from \ttbar\  production in association with other heavy-flavour quarks ($c{\bar c}$ or $b{\bar b}$) was absorbed into \cb.
\medskip

The largest systematic uncertainties in these measurements came from \ttbar\ modelling, PDFs, and imperfect knowledge of the integrated luminosities.
A summary of the uncertainties in \xtt\ is shown in
Table~\ref{t:systsum}. The uncertainty from each source was evaluated by repeatedly solving Eqs.~(\ref{e:tags}) with all their parameters simultaneously changed according to the parameter's dependence on a variation of $\pm 1$ standard deviation ($\sigma$) of that particular source. Correlated effects of the parameters variations on the measurements were thus taken into account. The total uncertainties in $\xtt$ and $\epsb$ were calculated by adding the statistical uncertainties and effects of all the individual systematic components in quadrature,
assuming them to be independent. In order to facilitate the combination with the CMS measurements, several ATLAS uncertainties are merged in quadrature compared to the list of uncertainties presented in Ref.~\cite{ATLASttbar1}. 
\medskip

The cross-sections were measured to be
\begin{eqnarray*}
  \xtt\ (\sxw ) & = & \ttxvalw\pm \ttxstatw ~\text{(stat.)} \pm\ttxsystw ~\text{(exp.+theo.)} \pm\ttxlumiw ~\text{(lumi.)}~\rm pb\   {\rm\ and}  \\
  \xtt\ (\sxv ) & = & \ttxval\pm\ttxstat ~\text{(stat.)}  \pm\ttxsyst ~\text{(exp.+theo.)} \pm\ttxlumi ~\text{(lumi.)}~\rm pb, 
\end{eqnarray*}
where the three uncertainties arose from the statistical power of the data, experimental and theoretical  systematic effects, and imperfect knowledge of the integrated luminosity, respectively. Although included in the original references, the LHC beam energy uncertainty is not used in the combination presented in this paper due to its reduced value as shown in Refs.~\cite{ATLASDiffXS, BeamE}.\footnote{This was a 1.8\% uncertainty in the \ttbar\ cross-section value for a 0.66\% uncertainty in beam energy, while the beam energy is now known to an accuracy of 0.1\%. }
The results are consistent with theoretical QCD calculations at NNLO+NNLL accuracy.

%
\begin{table}[htp]
 \caption{\label{t:systsum}Summary of the relative statistical, systematic
 and total uncertainties in the ATLAS measurements of the \ttbar\ production
 cross-section, \xtt, at $\sqrt{s}=7$ and 8~\TeV\ from Ref.~\cite{ATLASttbar1}, where ID stands for identification and JES for jet-energy scale. The grouping of the systematic uncertainties is modified with respect to Ref.~\cite{ATLASttbar1} to allow for the combination with the CMS measurements. }
 \vspace{0.2cm}
 \centering 
 \begin{tabular}{lcc}
   \hline\noalign{\smallskip}
ATLAS & \multicolumn{2}{c}{Merged uncertainty [\%]}  \\
Source                            &  7~\TeV  &  8~\TeV  \\
   \noalign{\smallskip}\hline\noalign{\smallskip}
   Trigger     &  0.2 &  0.2 \\
   Lepton (mis-)ID/isolation   &  0.9  &  0.8   \\
   Lepton energy scale &  0.3  &  0.5   \\
   JES flavour composition/specific response &  0.2 &  0.4 \\
   JES modelling                 &  ~~0.04 &  0.2 \\
   JES central/forward balance &  ~~0.03 &  0.1 \\
   JES pile-up  & ~~0.03 & 0.2 \\
   Other JES  & ~~0.03 & 0.2 \\
   Jet energy resolution  & 0.3 & 0.5 \\
   $b$-jet ID           &  0.4 &  0.4 \\
   $b$-jet mis-ID        &  ~~0.02 &  ~~0.02 \\
   \hline 
   $tW$ background &  0.8 &  0.8 \\
   Drell--Yan background &  ~~0.05 &  ~~0.02 \\
   Diboson background &  0.1 &  0.1 \\
   \hline
   \ttbar\ scale choice & 0.3 &  0.3 \\
   \ttbar\ generator modelling &  1.4 &  1.2 \\
   PDF &  1.0 &  1.1 \\
   \hline
   Integrated luminosity &  2.0 &  2.1 \\
  Statistical &  1.7 &  0.7 \\
   \hline
   {\bf Total uncertainty} &  {\bf 3.5} &  {\bf 3.2} \\
   \hline\noalign{\smallskip}
 \end{tabular}
\end{table}

\clearpage

\section{CMS measurements}
\label{sec:cms_meas}

The CMS measurements of \xtt at $\sqrt{s}=7$ and 8~\TeV\ were based on a binned likelihood fit to final-state observables and were also performed in the $e\mu$ channel~\cite{CMSttbar}. All events with an oppositely charged pair formed from an isolated electron and isolated muon were divided into three categories in $b$-tagged jet multiplicity ($N_b$). The event counts in the categories with exactly one or two $b$-tagged jets were expressed using Eq.~(\ref{e:tags}). The remaining events were assigned to a category with either zero or more than two $b$-tagged jets ($N_{0,\geq 3}$), for which the event count was given by

\begin{equation}
\label{eq:addcmscat}
N_{0,\geq 3} = L \xtt\ \epsem  - (N_1-N_1^\text{bkg}) - (N_2-N_2^\text{bkg}) +N_{0,\geq 3}^\text{bkg} \text{,}
\end{equation}
where $N_{0,\geq 3}^\text{bkg}$ is the number of background events in the category with either zero or more than two $b$-tagged jets.
For $b$-tagging, a high-purity working point with a 0.1\% average misidentification rate for light-flavour quark and gluon jets was used, such that the contribution from events with three or more $b$-tagged jets in this category is negligible.
The variables $\epsem$, $\cb$, and $\epsb$ in Eqs.~(\ref{e:tags}) and~(\ref{eq:addcmscat}) were centred at the values predicted by the simulation and varied according to the uncertainties assigned to the simulation as opposed to determining $\epsb$  and \xtt by solving the equations as is done for the ATLAS measurement.
\medskip

In each $N_b$ category, the events were further categorised according to the number of additional non-$b$-tagged jets. 
The cross-section extraction was performed treating all systematic uncertainties as nuisance parameters in a binned profile likelihood fit. As input to this fit, the {\pt}  distribution of the lowest-{\pt} additional jet was used, if present, to tighten the constraints on jet-energy scale (JES) uncertainties. In subcategories with zero additional jets, only the total event yield was fitted. The cross-sections were determined simultaneously at $\sqrt{s}=7$ and 8~\TeV. For this purpose, systematic uncertainties partially correlated between $\sqrt{s}=7$ and 8~\TeV\ were split into a correlated component and two uncorrelated contributions, one for each centre-of-mass energy.
Uncertainties were only constrained in the fiducial phase space defined by the kinematic acceptance for the $e\mu$ pair. Additional uncertainties were assigned to the extrapolation to the full phase space and were added in quadrature.
The largest contributions to the total uncertainty stem from trigger and lepton efficiencies, the Drell--Yan background modelling, and imperfect knowledge of the luminosity.
\medskip

The final results, assuming a top-quark mass of 172.5~\GeV, were
\begin{eqnarray*}
  \xtt\ (\sxw )  & = &
 173.6 \pm 2.1~\text{(stat.)}   ^{+4.5}_{-4.0}~\text{(exp.+theo.)}  \pm 3.8 ~\text{(lumi.)}  \text{ pb and}
\\
  \xtt\ (\sxv )& = &
 244.9 \pm 1.4~\text{(stat.)}   ^{+6.3}_{-5.5}~\text{(exp.+theo.)} \pm 6.4 ~\text{(lumi.)} \text{ pb,}
\end{eqnarray*}
where the uncertainties arose from the statistical power of the data, experimental and theoretical systematic effects, and imperfect knowledge of the integrated luminosity, respectively. As a result of the combined fit of the nuisance parameters and cross-sections, correlations were introduced among all fitted parameters, described by a covariance matrix which included the uncertainties in the nuisance parameters as well as the uncertainties in the cross-sections. For illustration, the impact of groups of related uncertainties is summarised in Table~\ref{t:cmssyssum}. 
The impact of each group of systematic uncertainties was estimated by fixing the corresponding parameters to their best-fit values, repeating the combination to assess the remaining uncertainty, and hence the size of an uncorrelated additional uncertainty that would reproduce the original total uncertainty. The latter estimate was taken as the impact of that uncertainty group, and only served as an illustrative estimate, since the full information is only contained in the full covariance matrix. For the statistical component, all nuisance parameters were fixed to their best-fit value and the remaining uncertainty contribution from statistics alone was evaluated. The results are in good agreement with theoretical QCD calculations at NNLO+NNLL accuracy.

\begin{table}[htp]
\caption{
Illustrative summary of the individual contributions to the total uncertainty in
the CMS \ttbar\ cross-section measurements from Ref.~\cite{CMSttbar}, where ID stands for identification, JES for jet-energy scale, ME for matrix element and PS for parton shower.
\label{t:cmssyssum}
}
\vspace{0.2cm}
\centering
\renewcommand{\arraystretch}{1.1}
\begin{tabular}{lcc}
\hline
CMS & \multicolumn{2}{c}{Uncertainty [\%]}  \\
Source                           &  7~\TeV  &  8~\TeV  \\
\hline
Trigger                     &  ${1.3}$ & ${1.2}$ \\
Lepton (mis-)ID/isolation         &  ${1.5}$ & ${1.5}$ \\
Lepton energy scale         &  ${0.2}$ & ${0.1}$ \\
JES total            &  ${0.8}$ & ${0.9}$\\
Jet energy resolution       &  ${0.1}$   & ${0.1}$ \\
$b$-jet ID                       &  ${0.5}$ &  ${0.5}$  \\
$b$-jet mis-ID                      &  ${0.2}$ &  ${0.1}$ \\
Pile-up                     &  ${0.3}$ &  ${0.3}$ \\
\hline
$tW$ background                    &  ${1.0}$ & ${0.6}$ \\
Drell--Yan background                          &  ${1.4}$ & ${1.3}$ \\
Non-$e\mu$ \ttbar              &  ${0.1}$ & ${0.1}$ \\
$\ttbar V$ background      &  ${0.1}$   & ${0.1}$ \\
Diboson  background         &  ${0.2}$ & ${0.6}$ \\
$W$+jets/QCD background      &  ${0.1}$   & ${0.2}$ \\
\hline
\ttbar\ scale choice                &  ${0.3}$ &  ${0.6}$ \\
ME/PS matching              &  ${0.1}$ &  ${0.1}$ \\
ME generator   &  ${0.4}$ &  ${0.5}$ \\
Hadronisation (JES)         &  ${0.7}$ &  ${0.7}$ \\
Top-quark \pt\ modelling                &  ${0.3}$ &  ${0.4}$ \\
Colour reconnection          &  ${0.1}$ &  ${0.2}$ \\
Underlying event            &  ${0.1}$   &  ${0.1}$ \\
PDF                         &  ${0.2}$ &  ${0.3}$ \\
\hline
Integrated luminosity                  &  ${2.2}$  &  ${2.6}$\\
Statistical                &  ${1.2}$ & ${0.6}$ \\
\hline  \\[-0.4cm]
\ttbar\ scale choice (extrapolation)    &  $^{+0.1}_{-0.4}$   & $^{+0.2}_{-0.1}$ \\
ME/PS matching (extrapolation)   & $^{+0.1}_{-0.1}$   & $^{+0.3}_{-0.3}$ \\
Top-quark \pt\ (extrapolation)      &  $^{+0.5}_{-0.3}$   & $^{+0.6}_{-0.3}$ \\
PDF (extrapolation)              &  $^{+0.1}_{-0.1}$   & $^{+0.1}_{-0.1}$ \\
\hline \\[-0.4cm]
{\bf Total uncertainty}                       &  {\bf $^{+3.6}_{-3.5}$  }& {\bf   $^{+3.7}_{-3.5}$ }\\[0.1cm]
\hline
\end{tabular}
\end{table}


\section{Combination method and assumptions}
\label{sec:combination}

 The CMS measurement was performed with simultaneously profiled uncertainties, leading to non-negligible post-fit correlations between them. 
Commonly used BLUE combination techniques~\cite{BLUEref1, BLUEref2} provide no method to account for these correlations, and it is known that neglecting these correlations can lead to severe biases and incorrect uncertainty estimates~\cite{Convino}. 
Therefore, the combination is performed using an algorithm which allows the consistent modelling of these correlations, implemented in the software tool Convino~\cite{Convino} using the covariance matrices of the individual measurements.
It is performed using a Pearson $\chi^2$~\cite{pearsonchi} minimisation, where the systematic uncertainties are represented by nuisance parameters. The $\chi^2$ is defined using three terms: one representing the results of each measurement and their statistical uncertainties, another one describing the correlations between the nuisance parameters and constraints on them from the data for each measurement, and finally a term incorporating prior knowledge of the systematic uncertainties and the assumed correlations between uncertainties, modelled by a multivariate Gaussian with non-zero correlations. The method takes as input the full covariance matrix provided by CMS, in addition to the publicly available data. For the ATLAS measurements, where the systematic uncertainties are unconstrained in the fit procedure, the uncertainties are either fully correlated or uncorrelated between $\sqrt{s}=7$ and 8~\TeV, and the correlated uncertainty sources are modelled by a single parameter per source.  
\medskip

The covariance matrix $C$ for a measurement can be expressed using four components

\begin{equation*}
C = 
\begin{pmatrix}
U & \kappa^T\\
\kappa & M
\end{pmatrix},
\end{equation*}
where the first diagonal block matrix $U$ describes the (co)variance of the nuisance parameters representing the systematic uncertainties. In the ATLAS case it is an identity matrix since the parameters describing the uncertainties are uncorrelated and are normalised to unity. A second block $M$ describes the measured cross-sections, and also has diagonal form, with entries representing the variance of the individual measurements. The last part of the matrix, denoted by $\kappa$, describes the impact of a 1$\sigma$ variation of a nuisance parameter on each of the cross-section measurements. 
The asymmetric extrapolation uncertainties in the CMS measurements have not been part of the fit to the data. These uncertainties are symmetrised by taking the maximum absolute impact as the symmetric uncertainty and are incorporated into the CMS covariance matrix using the same procedure. The components of the covariance matrix $C$ as well as the covariance matrix of the CMS input measurement are available in HEPdata~\cite{Hepdata}.

In this representation, the combination can be performed accounting for the correlation between the measurements as well as for the correlations and constraints within each individual measurement.
For this purpose, the constraints from the data are separated from the prior assumptions on the nuisance parameters within Convino.  These prior assumptions are assumed to follow a Gaussian distribution and express the range of a 1$\sigma$ variation of the respective parameters within the profile likelihood fit. After separating these terms, the individual $\chi^2$ terms of the  ATLAS and CMS measurements are combined. The Gaussian terms are reintroduced through a covariance matrix in which the assumed correlations between the systematic uncertainties from the ATLAS and CMS measurements are also included. A more detailed description of the method, including a validation based on pseudo-experiments, can be found in Ref.~\cite{Convino}.

\subsection{ATLAS and CMS systematic uncertainties}

The differences in the sizes of the systematic uncertainties between the input ATLAS and CMS measurements (Tables~\ref{t:systsum} and~\ref{t:cmssyssum}) arise from differences in the analysis methods and in the trigger and event selections. Full descriptions of each of the systematic uncertainties used in the individual measurements are available in Refs.~\cite{ATLASttbar1, CMSttbar}. As mentioned above, in order to facilitate the combination with the CMS measurements, several of the original ATLAS uncertainties are merged through summation in quadrature, while taking into account the correlations between the ATLAS $\sqrt{s}=7$ and~8~\TeV\ uncertainties. Given that all CMS uncertainties are all to some degree correlated, it is not possible to sum them into groups by adding them in quadrature. Instead, each component, or set of components has to be correlated with the corresponding merged uncertainty in the ATLAS measurement as described in the following section. 




\subsection{Correlation assumptions}
Several systematic uncertainties are assumed to be correlated between the ATLAS and CMS measurements. The correlation assumptions used in this combination are summarised in Table~\ref{t:assump}. For each correlated systematic uncertainty, except that in the integrated luminosity, the level of correlation is set to be one of the following values: LOW (0.25), HALF (0.5), HIGH (0.75) and FULL (1.0). Moreover, the correlations have also been scanned, one at a time, in a range around the given assumption value, bounded by the adjacent levels of correlation, e.g., two parameters correlated with a correlation coefficient of 0.5 are scanned from 0.25 to 0.75. No such variation changes the results significantly. The largest change originates from the luminosity correlation coefficient, which changes the total uncertainty by only 0.15\% when increased from 0.1 to 0.25. In cases where the sign of the correlation could not be determined unambiguously, the sign that maximises the total uncertainty of the combined result is chosen. This is the case for the correlation of, e.g., the matrix-element generator uncertainty of the CMS measurement with the generator uncertainty of the ATLAS result. Therefore, a conservative estimate is used in these cases.
\medskip

\begin{table}[htp]
  \caption{\label{t:assump}Assumed correlations between ATLAS and CMS systematic uncertainties. The assigned sign is based on the nature of the systematic uncertainty (e.g., minus for an `up' variation in the ATLAS measurement that corresponds to a `down' variation in the CMS measurement due to conventions within the collaborations). If the sign is ambiguous, the sign maximising the total uncertainty in the combined cross-section is chosen. Any uncertainties not included in the table are considered uncorrelated.}
  \vspace{0.2cm}
  \centering 
  \begin{tabular}{lrl}
    \hline
    ATLAS merged uncertainties      & Value&   CMS uncertainties \\
    \hline
          Lepton ID and energy resolution                 &  HALF                            &  Lepton ID and energy  resolution                                \\
    \hline
                            &  HIGH                            &  JES flavour  composition                                         \\
     JES flavour composition/specific response                   & $-$LOW                                           &  $b$-jet fragmentation tune                                        \\
             & LOW                                               &  $b$-jet neutrino decay fraction                                         \\
    \hline
     \multirow{2}{*}{JES modelling}  &  HALF                                       &  JES: AbsoluteMPFBias 7~\TeV                          \\
       &  HALF                                      &  JES: AbsoluteMPFBias 8~\TeV                          \\
    \hline
     \multirow{2}{*}{JES central/forward balance}      &  HIGH   &  JES: RelativeFSR 7~\TeV                              \\
           &  HIGH                                               &  JES: RelativeFSR 8~\TeV                              \\
    \hline
     \multirow{3}{*}{$tW$ background}        &  HIGH                &  $tW$ single top quark correlated         \\
             &  LOW                                              &  $tW$ single top quark 7~\TeV   \\
             &  LOW                                           &  $tW$ single top quark 8~\TeV   \\
    \hline
     \multirow{3}{*}{Diboson}         &  HIGH                   &  Diboson correlated                                                \\
              &  LOW                                             &  Diboson 7~\TeV                                        \\
              &  LOW                                            &  Diboson 8~\TeV                                        \\
    \hline
     \multirow{2}{*}{\ttbar\ scale choice}      &  HALF       &  \ttbar\ scale choice                                              \\
           &  HALF                                            &  \ttbar\  scale choice (extrapolation)                                          \\
    \hline
     \multirow{7}{*}{\ttbar\ generator}       & LOW           &  Top-quark \pt                                                   \\
             & LOW                                               &  Top-quark \pt\ (extrapolation)                                           \\
             & $-$LOW                                             &  ME generator                                          \\
             & LOW                                            &  ME/PS matching                                             \\
             & LOW                                            &  ME/PS matching (extrapolation)                                        \\
             & $-$LOW                                          &  Colour reconnection                                                      \\
             & $-$LOW                                           &  Underlying-event tune                                                      \\
    \hline
     Each PDF CT10 eigenvector    & FULL                                                &  Each PDF  CT10 eigenvector       

                                            \\ \hline
     Integrated luminosity    &  0.1                           & Integrated luminosity                                                \\

\hline
  \end{tabular}
\end{table}

{\bf Trigger:}
ATLAS uses single-lepton triggers, while CMS only considers dilepton triggers ($e\mu$). In addition, the techniques used to derive the trigger efficiencies are very different: in the case of ATLAS, the single-lepton trigger efficiencies are measured using tag-and-probe techniques, while for CMS, the efficiencies are derived from an orthogonal set of trigger paths based on missing transverse momentum. Therefore, the trigger uncertainties are considered to be uncorrelated.
\medskip

{\bf Lepton-related uncertainties:}
For both the ATLAS and CMS measurements, the lepton energy scale, resolution, identification and isolation efficiencies are studied using $Z\rightarrow ee/\mu\mu$, ${\rm{J}}/\Psi \rightarrow ee/\mu\mu$, and $W\rightarrow e \nu$ events. Parts of the corresponding uncertainties are detector- and algorithm-specific; however, other parts are systematic uncertainties related to the measurement methods, which are similar for the two experiments. Therefore, the correlation is considered to be HALF for the lepton-related uncertainties.
\medskip

{\bf Jet-energy scale:}
The correlations between the JES uncertainties from ATLAS and CMS follow the guidelines explained in Refs.~\cite{ATL-PHYS-PUB-2014-020, CombJES2}. Many components of the uncertainties, such as statistical and detector-related uncertainties from in situ techniques, pile-up uncertainties, high-\pt\ uncertainties, and jet fragmentation energy scale uncertainties, are considered uncorrelated, including the ATLAS $b$-jet energy scale uncertainty. Three remaining components are taken as partially correlated. The first component refers to the JES flavour composition uncertainties. These account for the flavour composition of the jets and the calorimeter response to jets of different flavours. In addition, for the CMS measurement, a variation of the $b$-hadron neutrino decay fraction as well as a variation of the $b$-quark fragmentation tune is also performed, which is assigned a LOW correlation with the flavour-dependent jet response in the ATLAS analysis. In this case, the sign of the chosen correlation coefficient cannot be determined unambiguously. Therefore, a negative correlation, which maximises the total uncertainty in the combined values, is chosen.
The second component includes the CMS {\it AbsoluteMPFBias}, which refers to the part of the absolute JES uncertainty related with the \pt-dependent calibration coming from a potential bias in the Missing Projection Fraction (MPF) method~\cite{CombJES2}. The corresponding merged uncertainty in ATLAS is estimated with a mix of the same MPF method and a method using \pt-balance between a jet and either a $Z$ or $\gamma$ reference object in the central $\eta$ region; this uncertainty is referred to as {\it JES modelling}. The recommended correlation between ATLAS and CMS from Refs.~\cite{ATL-PHYS-PUB-2014-020, CombJES2} for this uncertainty is between 0 and 0.5, therefore the correlation assigned is HALF. 

The third component includes the CMS {\it RelativeFSR}, which refers to the part of the relative JES uncertainty related to the $\eta$-dependent calibration and comes from the modelling of the final-state radiation effects. ATLAS refers to this transferring of the calibration to the forward region as {\it JES central/forward balance} (also referred to as the {\it $\eta$-intercalibration modelling} in some references).  The recommended correlation between ATLAS and CMS from Refs.~\cite{ATL-PHYS-PUB-2014-020, CombJES2} for this uncertainty is between 0.5 and 1.0. This component is considered uncorrelated between $\sqrt{s}=7$ and 8~\TeV\ in the case of the CMS measurements, and it is considered correlated between $\sqrt{s}=7$ and 8~\TeV\ for the ATLAS measurements. Because of this inconsistency between the two experiments, the highest correlation which does not lead to a non-positive-definite covariance matrix is 0.7 (HIGH), which is used in this combination. The effect of this choice is negligible since the contribution from RelativeFSR uncertainties is below 5\% of the total uncertainty for each input measurement.
\medskip

{\bf $b$-jet identification:}
The $b$-jet identification uncertainties also include uncertainties due to $c$-jet, light-flavour jet and gluon jet identification. In the ATLAS measurement, the $b$-jet identification efficiencies for signal events are determined in situ by solving Eq.~(\ref{e:tags}), consequently uncertainties affect the background contributions and the $b$-tagging correlation \cb, both derived from simulation. In the CMS measurement, the $b$-jet identification efficiencies are corrected by comparing efficiencies in data and simulation using a combination of several independent methods and calibration samples~\cite{CMS-PAS-BTV-13-001}. Therefore, these uncertainties are considered to be uncorrelated.
\medskip

{\bf Backgrounds:} The ATLAS $tW$ background uncertainty is the combination of four separate uncertainties assessed in the original ATLAS measurement: an uncertainty in the $tW$ cross-section, an uncertainty associated with the scheme handling the $tW$ and \ttbar\ interference, an uncertainty from the impact of initial- and final-state radiation on the $tW$ background, and an uncertainty from the generator used to simulate the $tW$ background. For both the $tW$ background and the diboson background, CMS includes an uncertainty component which is correlated between the measurements at $\sqrt{s}=7$ and 8~\TeV\ since Monte Carlo (MC) simulations are used to estimate the uncertainties from these backgrounds. This component is assigned a HIGH correlation with the corresponding ATLAS uncertainty because both experiments use a method relying on simulation. In addition, an uncertainty component specific to each centre-of-mass energy is used in the CMS measurement, for example related to using different MC parameters and parton shower tunings in the event generation for those backgrounds. This latter component is assigned a LOW correlation with the corresponding ATLAS uncertainty. Finally, due to the different event selections employed by ATLAS and CMS, the contributions from the Drell--Yan background are also different and in both analyses partly constrained by data control regions. The uncertainties associated with this background are considered as uncorrelated between ATLAS and CMS.
\medskip

{\bf \ttbar\ modelling:} The scale uncertainty refers to the uncertainties estimated by varying the factorisation and renormalisation scales in the \ttbar\ signal simulation by a factor of two. ATLAS uses the \textsc{Powheg\,Box}~\cite{Nason_2004, Frixione_2007, Frixione_2007_2} generator where the default scale is defined using the top-quark mass and transverse momentum as $Q^2 = m^2_t + p^2_{\mathrm{T},t}$ while CMS uses the \textsc{MadGraph}~\cite{Alwall_2014} generator where the default scale is $Q^2 = m^2_t + \Sigma p^2_{\mathrm{T}}$, where the sum runs over all additional final-state partons in the matrix-element calculations. Due to this slightly different $Q^2$ definition and also due to the fact that ATLAS varies the scales independently while CMS varies them simultaneously, the correlation is assumed to be HALF. The \ttbar\ generator uncertainty corresponds in the ATLAS case to comparing a sample of \ttbar\ events generated with \textsc{Powheg\,Box} interfaced to \textsc{Pythia\,6}~\cite{Sjostrand:2006za} with a sample generated with MC@NLO~\cite{Frixione_2002, Frixione_2003} interfaced to \textsc{Herwig}~\cite{Corcella_2001}, thereby incorporating different matrix-element treatments as well as different parton shower and hadronisation modelling. In the CMS case, the equivalent uncertainty was calculated by varying relevant parameters within the \textsc{MadGraph} generator.  In addition to affecting the fit of the cross-section in the fiducial phase space, these variations are applied a second time when extrapolating to the full phase space.
Since both methods address similar physical effects it is expected that the \ttbar\ modelling uncertainties will be somewhat correlated; however, since the methods and generators used to obtain them are different, the correlation is assumed to be LOW. Another contribution to the modelling uncertainties is the variation of the underlying-event and colour-reconnection tunes, as well as a comparison between different matrix-element generators and the reweighting of the simulated top-quark \pt\ spectrum to match the one observed in data in the case of the CMS measurement. These individual sources are assigned a LOW correlation with the combined ATLAS generator uncertainty. Also, here the sign of some components is chosen such that the total uncertainty is maximised.
\medskip

{\bf PDF:} For the purpose of the combination, the ATLAS \ttbar\ PDF uncertainty is split into two components: an uncertainty from the CT10 eigenvectors only, which is considered fully correlated with the CMS PDF CT10 uncertainties, and a remainder PDF uncertainty (from MSTW2008 and NNPDF2.3) which is uncorrelated with the CMS PDF uncertainty. The sum in quadrature of these two ATLAS components equals the original ATLAS PDF uncertainty.
\medskip

{\bf Integrated luminosity:} The integrated-luminosity uncertainty affects the determination of the signal yield and most background yields. This uncertainty is assumed to be uncorrelated between the $\sqrt{s}=7$ and 8~\TeV\ data by both the ATLAS and CMS measurements as well as in the combination. For the $\sqrt{s}=7~\TeV$ (8~\TeV) running period, the luminosity uncertainty for ATLAS is 1.8\% (1.9\%)~\cite{ATLAS_lumi, ATLAS_lumi2}, with 1.5\% (1.2\%) estimated from the van der Meer scan analysis and 0.9\% (1.5\%) from the long-term luminosity monitoring and the transfer of the luminosity scale from the van der Meer regime to the physics regime. For CMS the uncertainties are 2.2\% (2.6\%) for 7~\TeV\  (8~\TeV)~\cite{CMS-PAS-SMP-12-008, CMS-PAS-LUM-13-001}, of which 1.8\% (2.3\%) is estimated from the van der Meer scan analysis, and 1.2\% (1.2\%) from the luminosity-monitoring uncertainty. The uncertainty estimated from the long-term luminosity monitoring is detector-specific and thus uncorrelated between ATLAS and CMS. The uncertainty estimated from the van der Meer scan analysis is partially correlated. For the $\sqrt{s}=7~\TeV$ (8~\TeV) running period, the correlated components amount to 0.5\% (0.6\%) and 0.5\% (0.7\%) for ATLAS and CMS, respectively; they arise from the measurement of the bunch intensities based on a common device, the correction for beam-to-beam-induced biases extracted from a common simulation, and the models used to fit the visible interaction rate as a function of the separation between the beams, referred to as beam modelling. Taking into account the covariance matrix built from those correlations, the resulting correlation coefficients are computed to be 0.065 for the 7~\TeV\ running period and 0.085 for the 8~\TeV\ running period. Both values are rounded to 0.1 for this combination.

\section{Results}
\label{sec:results_and_mass}

\subsection{Cross-section combination}
\label{subsec:xsection}

\begin{figure}[htbp]
\centering
\includegraphics[width=0.595\textwidth]{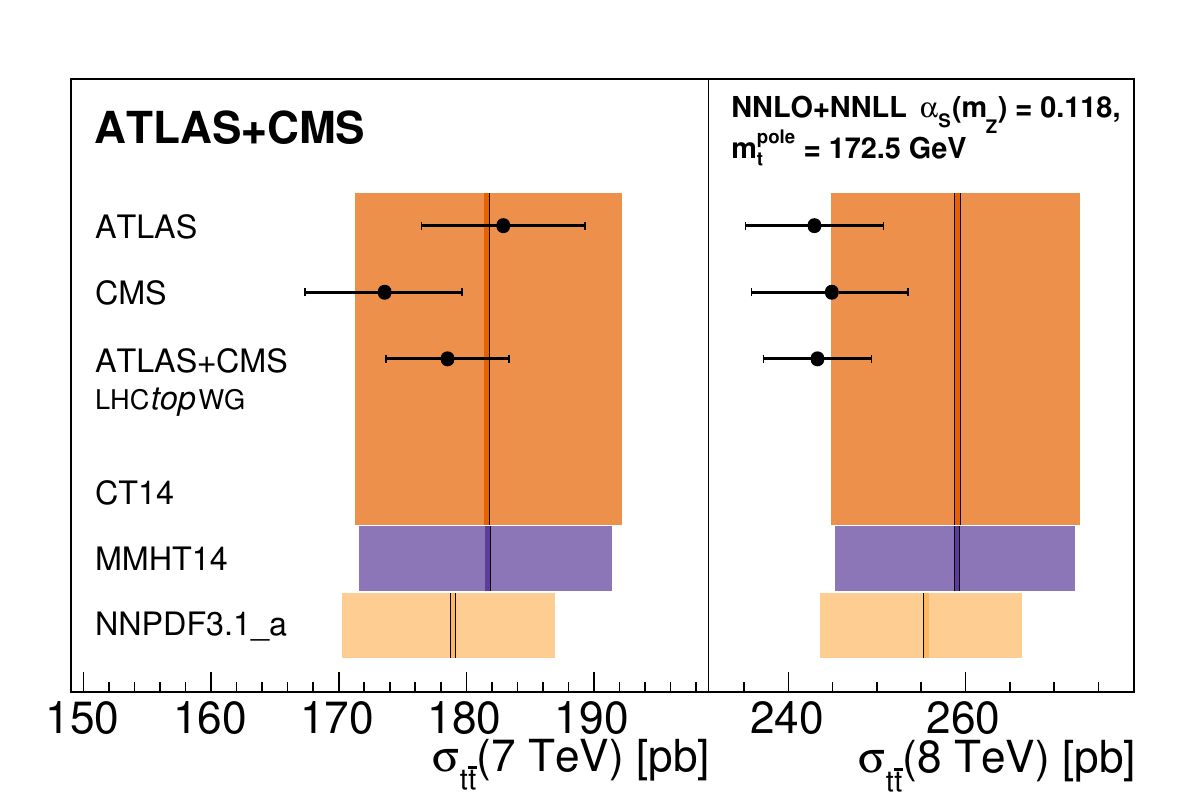}
\caption{Measured \ttbar\ production cross-sections at $\sqrt{s}=7$ and~8~\TeV\ compared with predictions using different PDF sets. NNPDF3.1\_a is a version of this PDF set containing no top-quark measurements. The shaded bands represent the total uncertainties in the predictions.
\label{fig:combination}}
\end{figure}

The combination of the cross-sections is performed simultaneously at $\sqrt{s}=7$ and 8~\TeV, such that the corresponding correlations are taken into account. The resulting cross-sections are
\begin{eqnarray*}
\sigma_{\ttbar}\ (\sxw )  & = &
 \ttxvalcombs  \pm \ttxtotpcombs  \mbox{ pb}
\\
\sigma_{\ttbar}\ (\sxv ) & = &
 \ttxvalcombe   ^{+\ttxtotpcombe}_{-\ttxtotmcombe}  \mbox{ pb}
\text{,}
\end{eqnarray*}
with a correlation between the $\sqrt{s}=7$ and 8~\TeV\ values of $\rho=\rhocomb$  and minimum Pearson $\chi^2$ of $1.6$ for two degrees of freedom.
A comparison of the combined result with the input measurements and the prediction using different PDF sets  is shown in Figure~\ref{fig:combination}.  
\medskip

The impact of individual groups of uncertainties on the combined results is estimated as is done for the CMS input measurement, described in Section~\ref{sec:cms_meas}. 
The resulting impacts are listed in Table~\ref{t:sysimpact}. The integrated luminosity is still the dominant uncertainty although its impact is reduced by up to 35\% compared to the individual results. The statistical uncertainty is reduced by up to 40\% by the combination. The next largest uncertainty in the ATLAS measurement was the \ttbar\ generator uncertainty, which is reduced by 40\% by the combination, while for CMS the next largest uncertainty was associated with the lepton ID and energy, and is also reduced by up to 40\% by the combination. Overall, the nuisance parameter constraints are similar to the ones coming from the CMS input measurement. The observed reduction of the luminosity uncertainty on the combined cross-section is consistent with the expected reduction factor for this weighted average, given the magnitude of the correlated and uncorrelated components of the ATLAS and CMS luminosity measurements. Relative to the most precise input measurements, the combination improves the precision by 25\% (28\%) at $\sqrt{s}=7$~\TeV\ (8~\TeV), and therefore the combined results are the most precise measurements of the inclusive \ttbar\ cross-section to date at those centre-of-mass energies. The experimental uncertainty is smaller than the theoretical uncertainty of the NNLO+NNLL predictions for the corresponding cross-sections.
\medskip


 
\begin{table}[htp]
  \caption{\label{t:sysimpact} Illustration of the impact $\Delta\sigma_{\ttbar}/\sigma_{\ttbar}$ of the dominant groups of systematic uncertainties on the combined cross-sections at $\sqrt{s}=7$ and~8~\TeV.}
  \vspace{0.2cm}
  \centering 
  \begin{tabular}{l c c}
  \hline
 Uncertainty & $\Delta\sigma_{\ttbar}(7~\TeV)$~[\%]        & $\Delta\sigma_{\ttbar}(8~\TeV)$~[\%]         \\
 \hline
Trigger                          &  0.6        & 0.5        \\
Lepton (mis-)ID, isolation and energy          & 1.0        & 0.9        \\
JES flavour composition &  0.4        &  0.4         \\
JES modelling                  &  $<0.1$~~~~        & 0.1         \\
JES central/forward balance                 &  0.2        & 0.2        \\ 
$b$-jet (mis-)ID                      & 0.4        & 0.4        \\ 
Pile-up                             & 0.2        & 0.2        \\ 
\hline
$tW$ background              & 0.8        & 0.6        \\
Drell--Yan background                        & 0.7         & 0.4       \\
Diboson background                    &  0.2        & 0.4       \\
\hline
\ttbar\ generator                 &   0.8       &  0.8     \\
\ttbar\ scale choice                       &    0.4        & 0.4       \\
PDF                                 & 0.4      &  0.3       \\ 
\hline
Integrated luminosity                            &   1.7        &  1.7       \\
Statistical                              &   1.0        & 0.4       \\  \hline \\[-0.4cm]
{\bf Total uncertainty} & {\bf $^{+2.7}_{-2.6}$} &{\bf   $^{+2.5}_{-2.4}$}\\[0.1cm]
\hline

  \end{tabular}
 \end{table}

The ratio of the cross-section at $\sqrt{s}=8$~\TeV\ to that at 7~\TeV\ is determined to be

\begin{equation}\nonumber
R_{8/7} = \ttxratio \pm \ttxratiostaterr ~\text{(stat.)} \pm\ttxratiosysterr ~\text{(syst.)} \text{,}
\end{equation}

based on the fitted values of the individual cross-sections, and accounting for the correlated uncertainties. 

\begin{figure}[htbp]
\centering
\includegraphics[width=0.695\textwidth]{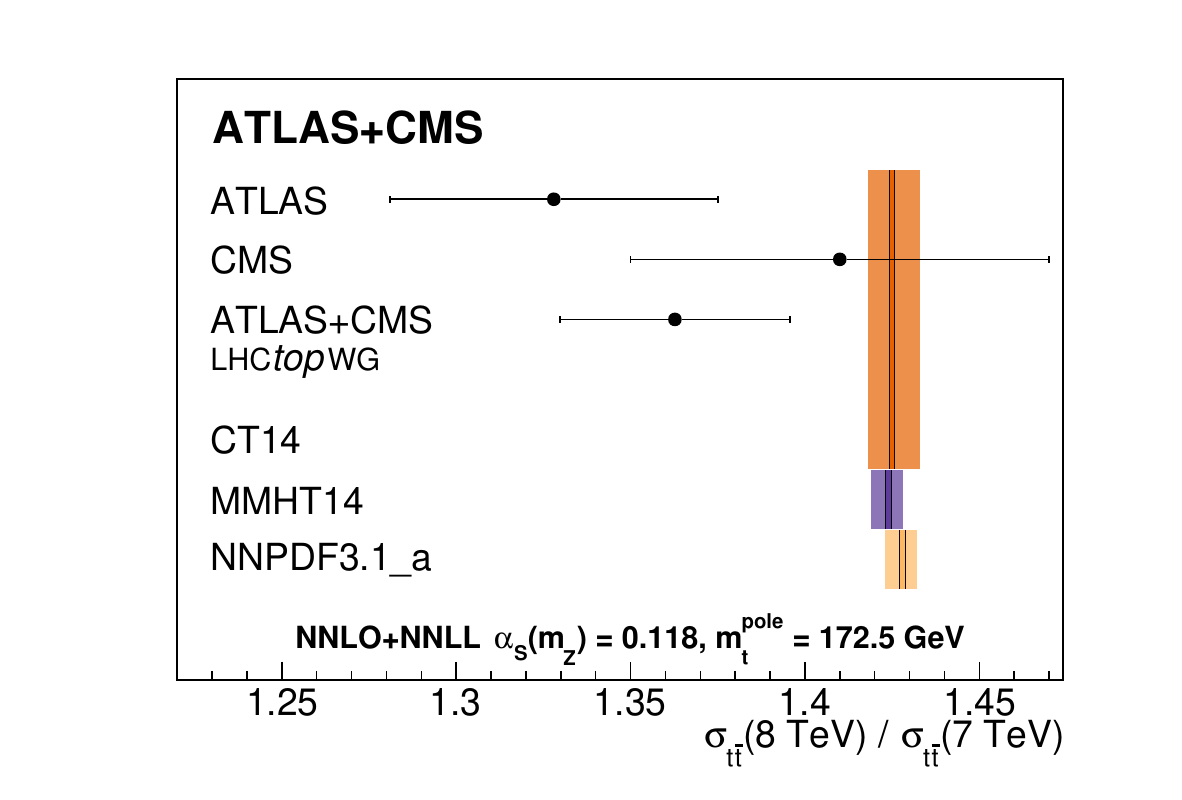}
\caption{Measured ratios of the \ttbar\ production cross-sections at $\sqrt{s}=7$ and~8~\TeV\ compared with predictions using different PDF sets. NNPDF3.1\_a is a version of this PDF set containing no top-quark measurements. The shaded bands represent the total uncertainties in the predictions.
\label{fig:ratio}}
\end{figure}

The measured ratio is compared with predictions using different PDF sets in Figure~\ref{fig:ratio}. Since correlated uncertainties cancel out in the ratio, uncorrelated sources, such as the statistical uncertainty, play a larger role. The cross-section ratio therefore benefits most from the combination, with the precision improving by 45\% relative to the most precise input measurements. Both the individual cross-sections and their ratio are in agreement with the SM prediction.  The level of agreement observed when comparing the measured 7~\TeV\ cross-section, the 8~\TeV\ cross-section and the ratio with the corresponding predictions is found to be 0.3$\sigma$, 1.0$\sigma$ and 1.9$\sigma$ respectively,
using $\mtpole=172.5$~\GeV, the CT14 PDF, and taking into account the uncertainties in the combined cross-sections and the predictions. 
The uncertainties in the predicted ratios include the effects on the cross-section calculations from the renormalisation and factorisation scale uncertainties (treated as correlated between $\sqrt{s}=7$ and 8~\TeV), the PDF uncertainty (the ratio is evaluated for each individual PDF eigenvector and the relevant prescription for the given PDF set is applied), $\alphas(m_Z) = 0.118 \pm 0.001$, and a top-quark pole mass uncertainty of 1.0~\GeV\ with 172.5~\GeV\ as the central value (both treated as fully correlated between $\sqrt{s}=7$ and 8~\TeV). The predicted ratio's uncertainty is smaller than the uncertainty in the measured ratio since those effects mostly cancel out in the ratio, and it is dominated by the PDF uncertainty, while variations of the pole mass and $\alphas(m_Z)$ do not contribute significantly to the predicted ratio's uncertainty (also described in Ref.~\cite{Mangano2012}).

\subsection{Top-quark pole mass and strong coupling}
\label{subsec:polemass}
The predicted value of the inclusive cross-section for \ttbar\ production is very sensitive to \mtpole\ and $\alphas(m_Z)$. The estimates of \mtpole and $\alphas(m_Z)$ from the inclusive cross-section measurement are fully correlated. Therefore, the measured cross-sections can only be used to extract a measurement of $\alphas(m_Z)$ by using an assumed value for \mtpole and vice versa.  Either parameter can be determined by comparing the combined cross-sections with their predictions as a function of that parameter while fixing the other. While a few percent change in $\alphas(m_Z)$ in the matrix-element calculation leads to a negligible change in the measured cross-section, the extrapolation to the full phase space has a mild residual dependence of the order of 0.2\% per~\GeV\ on the top-quark mass for both the ATLAS and CMS measurements. The parameterisation of this dependence is different for each experiment, although the dependence is almost identical for \mtpole\ between 170 and 180~\GeV. To account for the different functional forms, the combination is performed assuming top-quark masses of 166.5~\GeV\ and 178.5~\GeV, in addition to the nominal value of 172.5~\GeV. For each of the three mass points, weights for the ATLAS and CMS measurements are determined based on their contribution to the combination at that mass point. The weights at other mass points are interpolated or extrapolated linearly from the values at the reference points. The final parameterisation used for the parameter extraction is calculated as a weighted mean of the ATLAS and CMS parameterisations, using the weights from this linear interpolation or extrapolation. The relative uncertainty in the combined results is also interpolated or extrapolated linearly over the mass range, and is almost constant. Finally, the ambiguities in the interpretation of the top-quark mass involved in MC and fixed-order calculations imply an additional uncertainty obtained by shifting the mass in the acceptance dependence by $\pm 1.0$~\GeV~\cite{TopMassHoang}. Due to the mild dependence of the measurements on this mass assumption, this contribution is negligible in the final result.
\medskip

\begin{figure}[ht]
\centering
\begin{subfigure}{7cm}
\centering
\includegraphics[width=7.5cm]{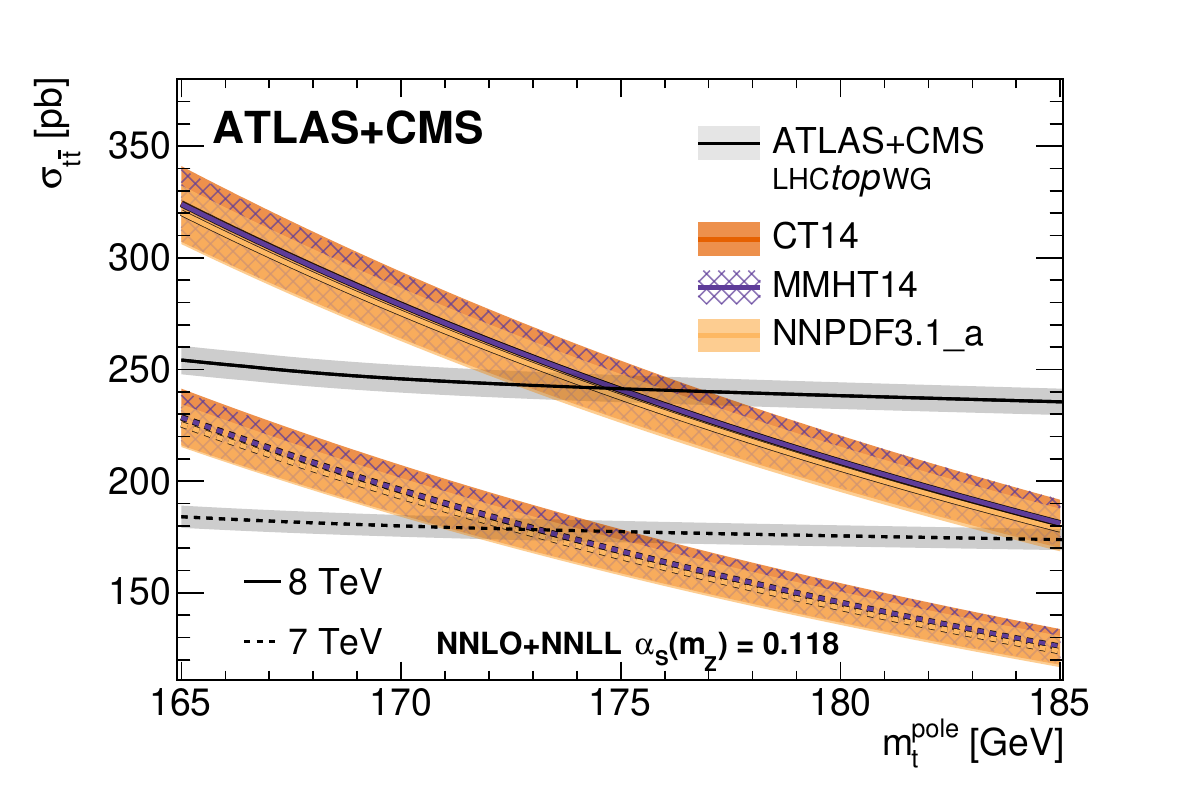}
\caption{}
\end{subfigure}
\begin{subfigure}{7cm}
\centering
\includegraphics[width=7.5cm]{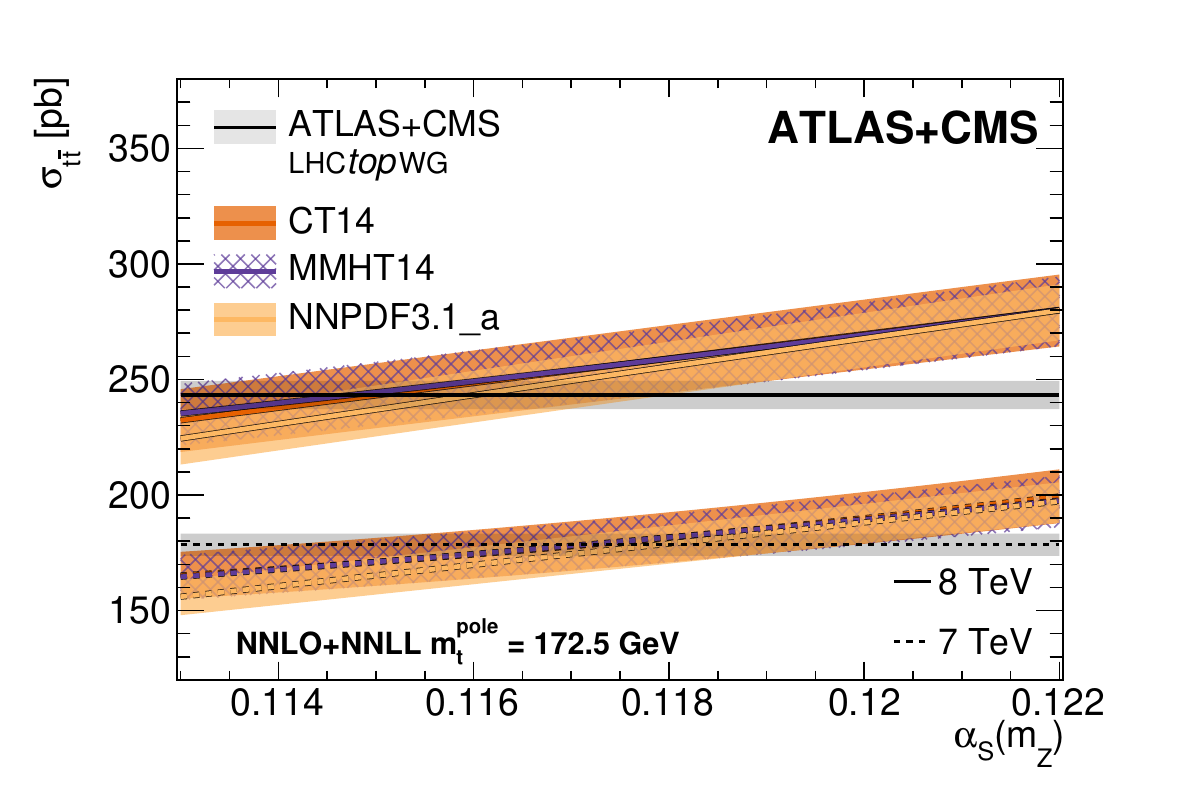}
\caption{}
\end{subfigure}
\caption{Dependence of the predicted cross-section and the combined measured cross-section on (a) the top-quark pole mass {\mtpole} and (b) the strong coupling $\alphas(m_Z)$. The prediction is evaluated for three different PDF sets and assumes (a) $\alphas(m_Z)=0.118 \pm 0.001$ or (b) $\mtpole = 172.5 \pm 1.0$~\GeV. NNPDF3.1\_a is a version of this PDF set containing no top-quark measurements. The uncertainty bands include the effects of the uncertainties in the combined cross-sections (interpolated or extrapolated linearly from the values at the reference points) and of the factorisation and renormalisation scale and PDF uncertainties in the predicted cross-sections.
\label{f:xsecdep}}
\end{figure}

The predicted dependence of the \ttbar\ cross-section on the top-quark pole mass is evaluated using \toppp, assuming $\alphas(m_Z)=0.118$ as a baseline. The dependence is derived using ten mass points. In addition, the prediction is evaluated for five different values of $\alphas(m_Z)$, assuming 
$\mtpole =172.5$~\GeV, by varying $\alphas(m_Z)$ consistently in the PDF and the calculation. In each case the cross-section is fitted with a fourth-order polynomial. The relative effects of these variations are assumed to be independent.
The comparisons with the combined cross-sections are shown in Figure~\ref{f:xsecdep}. The relative variations with $\mtpole$ and $\alphas(m_Z)$ are very similar for different PDF sets and the mild dependence of the measured cross-sections on $\mtpole$ is visible.
\medskip

\begin{figure}[htbp]
\centering
\includegraphics[width=0.495\textwidth]{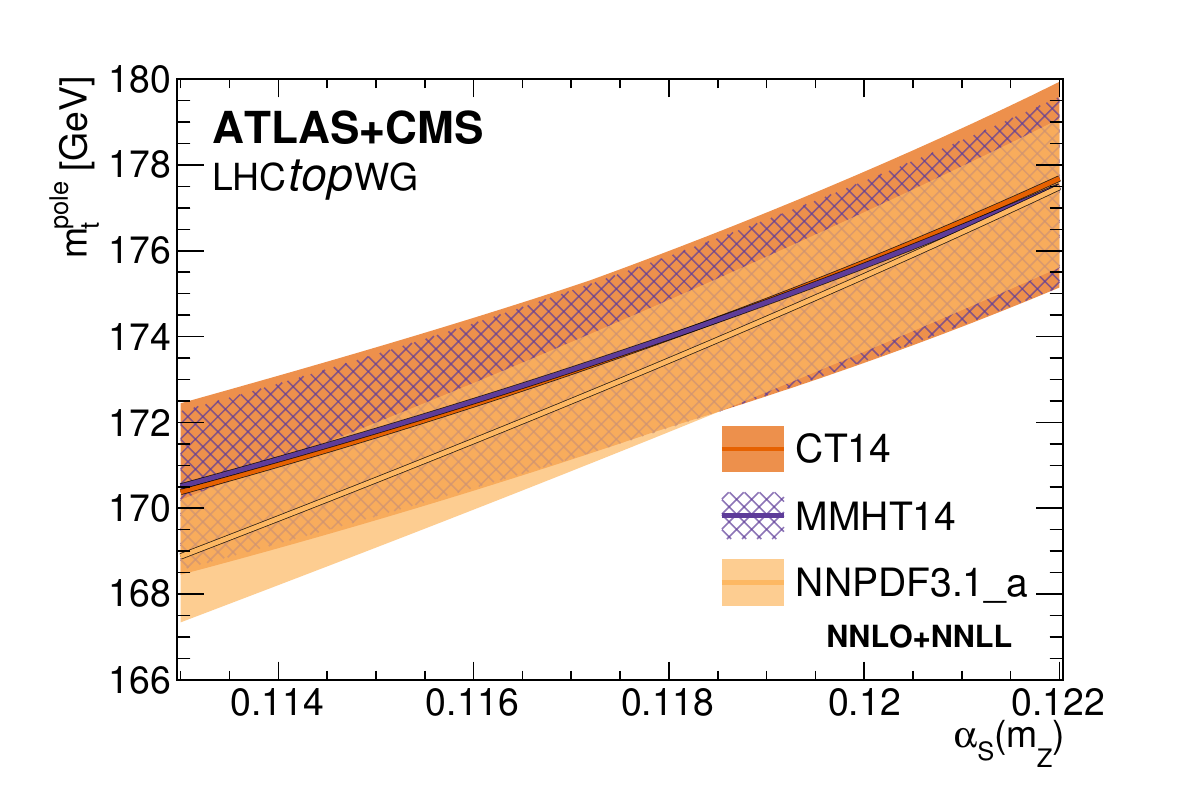}
\caption{Dependence of the top-quark pole mass, \mtpole, on the assumed value for the strong coupling, $\alphas(m_Z)$, including constraints from the combined measured cross-sections for three different PDF sets. NNPDF3.1\_a is a version of this PDF set containing no top-quark measurements. The shaded band represents the 68\% CL uncertainty on the extracted $\alphas(m_Z)$ when fixing \mtpole and on \mtpole when fixing $\alphas(m_Z)$. The uncertainty bands include the effects of the uncertainties in the combined cross-sections and of the renormalisation and factorisation scale and PDF uncertainties in the predicted cross-sections.
\label{fig:asmt}}
\end{figure}

The values of $\alphas(m_Z)$ and \mtpole\ are extracted from the combined cross-sections using a $\chi^2$ minimisation technique. The $\chi^2$ is defined as 
\begin{equation}
\chi^2 = \frac{1}{1-\rho^2}   \left(
\Delta(7~\TeV)^2 + \Delta(8~\TeV)^2 - 2 \rho \Delta(7~\TeV)\Delta(8~\TeV) \right) \text{, with }
\label{eq:chi2}
\end{equation} 
\begin{equation*}
\Delta = \frac{\xtt(\mtpole) - \xtt^\text{p}(\mtpole, \alphas(m_Z))}{\delta} \text{,}
\label{eq:delta}
\end{equation*}
and where $\xtt^\text{p}$ is the predicted \ttbar\ cross-section as a function of \mtpole and $\alphas(m_Z)$,  $\xtt(\mtpole)$ is the measured cross-section with a residual pole mass dependence, $\delta$ represents the experimental uncertainty of the combined cross-section, and $\rho$ is the correlation coefficient (\rhocomb) between the combined cross-section values. The uncertainties in the \ttbar\ cross-section prediction from each renormalisation and factorisation scale and PDF eigenvector variation are propagated to the final result by re-extracting the top-quark pole mass with different assumptions about the scales and PDF. The total scale uncertainty is determined from the envelope of its individual contributions, and the total PDF uncertainty is calculated using the prescription of the corresponding PDF set. For the case of the CT14 PDF set, the uncertainties are rescaled to 68\% CL, as appropriate for this PDF set. Figure~\ref{fig:asmt} shows the constraints in the $\alphas(m_Z)$--\mtpole plane from the combined cross-sections. The uncertainty bands include the effects of the uncertainties in the combined cross-sections and of the renormalisation and factorisation scale and PDF uncertainties in the predicted cross-sections.
\medskip

The dependence shown in Figure~\ref{fig:asmt} is used subsequently to extract either \mtpole or $\alphas(m_Z)$. The confidence intervals cover the negligible difference that results from minimising Eq.~(\ref{eq:chi2}) with respect to \mtpole while stepping over values of \alphas\ instead of minimising with respect to \alphas\ while stepping over values of \mtpole. Therefore,  Figure~\ref{fig:asmt}  can be used to extract $\alphas(m_Z)$ for a fixed value of \mtpole and vice versa.
When extracting \mtpole,  $\alphas(m_Z)=0.118 \pm 0.001$ is used, whereas when determining $\alphas(m_Z)$, \mtpole is assumed to be $\mtpole=172.5 \pm 1.0$~\GeV. The corresponding results for each PDF set are reported in Table~\ref{t:massalphares}. These results represent the most precise determinations of $\alphas(m_Z)$ at the scale of inclusive \ttbar\ production ($\mathcal{O}(2 \mtpole)$), being more precise than the previous measurements reported by CMS in Refs.~\cite{firstttbaralphas, Sirunyan_2019}, and are among the most precise top-quark pole mass measurements. The results for \mtpole\ are compatible with the values reported in Refs.~\cite{ATLASttbar1, CMSttbar}. 

\begin{table}[htp]
  \caption{\label{t:massalphares} Measured \mtpole\ and $\alphas(m_Z)$ values for each PDF set using the measured 7 and 8~\TeV\ combined cross-sections. NNPDF3.1\_a is a version of this PDF set containing no top-quark measurements.}
  \vspace{0.2cm}
  \centering 
  \renewcommand{\arraystretch}{1.4}
   \begin{tabular}{lcc}
    \hline
PDF set & \mtpole\  & $\alphas(m_Z)$ \\ 
              & ($\alphas= 0.118\pm 0.001$) & ($m_t=172.5\pm 1.0~\GeV$) \\ \hline
CT14 & \massCT\ $^{+\massuncCTup}_{-\massuncCTdown}$~\GeV\ & \alphasCT\ $^{+\alphasuncCTup}_{-\alphasuncCTdown}$ \\
MMHT2014 & \massMMHT\ $^{+\massuncMMHTup}_{-\massuncMMHTdown}$~\GeV\ & \alphasMMHT\ $^{+\alphasuncMMHTup}_{-\alphasuncMMHTdown}$ \\
NNPDF3.1\_a & \massNNascorrnotop\ $^{+\massuncNNascorrnotopup}_{-\massuncNNascorrnotopdown}$~\GeV\ & \alphasNNascorrnotop\ $^{+\alphasuncNNascorrnotopup}_{-\alphasuncNNascorrnotopdown}$ \\
\hline
  \end{tabular}
  \renewcommand{\arraystretch}{1.0}
\end{table}



\FloatBarrier
\section{Summary and conclusions}
\label{sec:conclusions}
A combination of measurements of the inclusive \ttbar\ production cross-section performed by the ATLAS and CMS experiments at $\sqrt{s}= 7$ and 8~\TeV\ is presented, accounting for correlations between the measurements from different experiments
as well as correlations within the ATLAS and CMS measurements. The resulting cross-sections are
\begin{eqnarray*}
  \xtt\ (\sxw )  & = &
 \ttxvalcombs  \pm \ttxtotpcombs  \mbox{ pb}
\\
  \xtt\ (\sxv )  & = &
 \ttxvalcombe   ^{+\ttxtotpcombe}_{-\ttxtotmcombe}  \mbox{ pb.}
\end{eqnarray*}

The combined results improve on the precision of the most precise individual results by 25\% at $\sqrt{s}=7$~\TeV, and by 28\% at $\sqrt{s}=8$~\TeV, making these combined results the most precise measurements of the inclusive \ttbar\ cross-section to date at those respective centre-of-mass energies.
The correlation between the combined cross-sections values is 0.41, and their ratio is determined to be
\begin{equation}\nonumber
 R_{8/7}=\ttxratio \pm \ttxratioerr \mbox{.}
 \end{equation}
 
Furthermore, the combined values for \xtt are used to determine the top-quark pole mass and the strong coupling by comparing  them with the predicted evolution of $\sigma_{\ttbar}$ as a function of \mtpole and $\alphas(m_Z)$, for different PDF sets. The measurement yields the most precise values, $\mtpole=  \massNNascorrnotop ^{+\massuncNNascorrnotopup}_{-\massuncNNascorrnotopdown}~\GeV$ (\massuncrelNNascorrnotop\ relative uncertainty) and $\alphas(m_Z) = \alphasNNascorrnotop^{+ \alphasuncNNascorrnotopup}_{-\alphasuncNNascorrnotopdown}$ (\alphasuncrelNNascorrnotop\ relative uncertainty), when using the NNPDF3.1\_a PDF set. The extracted $\alphas(m_Z)$ value is more precise than previous measurements performed using top-quark events.


%% file: Acknowledgements_ATLAS.tex


ATLAS thanks CERN for the very successful operation of the LHC, as well as the
support staff from our institutions without whom ATLAS could not be
operated efficiently.

ATLAS acknowledges the support of 
ANPCyT, Argentina; 
YerPhI, Armenia; 
ARC, Australia; 
BMWFW and FWF, Austria; 
ANAS, Azerbaijan; 
CNPq and FAPESP, Brazil; 
NSERC, NRC and CFI, Canada; 
CERN; 
ANID, Chile; 
CAS, MOST and NSFC, China; 
Minciencias, Colombia; 
MEYS CR, Czech Republic; 
DNRF and DNSRC, Denmark; 
IN2P3-CNRS and CEA-DRF/IRFU, France; 
SRNSFG, Georgia; 
BMBF, HGF and MPG, Germany; 
GSRI, Greece; 
RGC and Hong Kong SAR, China; 
ISF and Benoziyo Center, Israel; 
INFN, Italy; 
MEXT and JSPS, Japan; 
CNRST, Morocco; 
NWO, Netherlands; 
RCN, Norway; 
MEiN, Poland; 
FCT, Portugal; 
MNE/IFA, Romania; 
MESTD, Serbia; 
MSSR, Slovakia; 
ARRS and MIZ\v{S}, Slovenia; 
DSI/NRF, South Africa; 
MICINN, Spain; 
SRC and Wallenberg Foundation, Sweden; 
SERI, SNSF and Cantons of Bern and Geneva, Switzerland; 
MOST, Taiwan; 
TENMAK, T\"urkiye; 
STFC, United Kingdom; 
DOE and NSF, United States of America. 
In addition, individual groups and members have received support from 
BCKDF, CANARIE, Compute Canada and CRC, Canada; 
PRIMUS 21/SCI/017 and UNCE SCI/013, Czech Republic; 
COST, ERC, ERDF, Horizon 2020 and Marie Sk{\l}odowska-Curie Actions, European Union; 
Investissements d'Avenir Labex, Investissements d'Avenir Idex and ANR, France; 
DFG and AvH Foundation, Germany; 
Herakleitos, Thales and Aristeia programmes co-financed by EU-ESF and the Greek NSRF, Greece; 
BSF-NSF and MINERVA, Israel; 
Norwegian Financial Mechanism 2014-2021, Norway; 
NCN and NAWA, Poland; 
La Caixa Banking Foundation, CERCA Programme Generalitat de Catalunya and PROMETEO and GenT Programmes Generalitat Valenciana, Spain; 
G\"{o}ran Gustafssons Stiftelse, Sweden; 
The Royal Society and Leverhulme Trust, United Kingdom.

The crucial computing support from all WLCG partners is acknowledged gratefully, in particular from CERN, the ATLAS Tier-1 facilities at TRIUMF (Canada), NDGF (Denmark, Norway, Sweden), CC-IN2P3 (France), KIT/GridKA (Germany), INFN-CNAF (Italy), NL-T1 (Netherlands), PIC (Spain), ASGC (Taiwan), RAL (UK) and BNL (USA), the Tier-2 facilities worldwide and large non-WLCG resource providers. Major contributors of computing resources are listed in Ref.~\cite{ATL-SOFT-PUB-2021-003}.


%% file: CMS-TOP-18-014-acks.tex
CMS congratulates colleagues in the CERN accelerator departments for the excellent performance of the LHC and thank the technical and administrative staffs at CERN and at other CMS institutes for their contributions to the success of the CMS effort. In addition, CMS gratefully acknowledges the computing centres and personnel of the Worldwide LHC Computing Grid and other centres for delivering so effectively the computing infrastructure essential to our analyses. Finally, CMS acknowledges the enduring support for the construction and operation of the LHC, the CMS detector, and the supporting computing infrastructure provided by the following funding agencies: BMBWF and FWF (Austria); FNRS and FWO (Belgium); CNPq, CAPES, FAPERJ, FAPERGS, and FAPESP (Brazil); MES and BNSF (Bulgaria); CERN; CAS, MoST, and NSFC (China); MINCIENCIAS (Colombia); MSES and CSF (Croatia); RIF (Cyprus); SENESCYT (Ecuador); MoER, ERC PUT and ERDF (Estonia); Academy of Finland, MEC, and HIP (Finland); CEA and CNRS/IN2P3 (France); BMBF, DFG, and HGF (Germany); GSRI (Greece); NKFIH (Hungary); DAE and DST (India); IPM (Iran); SFI (Ireland); INFN (Italy); MSIP and NRF (Republic of Korea); MES (Latvia); LAS (Lithuania); MOE and UM (Malaysia); BUAP, CINVESTAV, CONACYT, LNS, SEP, and UASLP-FAI (Mexico); MOS (Montenegro); MBIE (New Zealand); PAEC (Pakistan); MES and NSC (Poland); FCT (Portugal);  MESTD (Serbia); MCIN/AEI and PCTI (Spain); MOSTR (Sri Lanka); Swiss Funding Agencies (Switzerland); MST (Taipei); MHESI and NSTDA (Thailand); TUBITAK and TENMAK (Turkey); NASU (Ukraine); STFC (United Kingdom); DOE and NSF (USA).

\hyphenation{Rachada-pisek} Individuals have received support from the Marie-Curie programme and the European Research Council and Horizon 2020 Grant, contract Nos.\ 675440, 724704, 752730, 758316, 765710, 824093, 884104, and COST Action CA16108 (European Union); the Leventis Foundation; the Alfred P.\ Sloan Foundation; the Alexander von Humboldt Foundation; the Belgian Federal Science Policy Office; the Fonds pour la Formation \`a la Recherche dans l'Industrie et dans l'Agriculture (FRIA-Belgium); the Agentschap voor Innovatie door Wetenschap en Technologie (IWT-Belgium); the F.R.S.-FNRS and FWO (Belgium) under the ``Excellence of Science -- EOS" -- be.h project n.\ 30820817; the Beijing Municipal Science \& Technology Commission, No. Z191100007219010; the Ministry of Education, Youth and Sports (MEYS) of the Czech Republic; the Hellenic Foundation for Research and Innovation (HFRI), Project Number 2288 (Greece); the Deutsche Forschungsgemeinschaft (DFG), under Germany's Excellence Strategy -- EXC 2121 ``Quantum Universe" -- 390833306, and under project number 400140256 - GRK2497; the Hungarian Academy of Sciences, the New National Excellence Program - \'UNKP, the NKFIH research grants K 124845, K 124850, K 128713, K 128786, K 129058, K 131991, K 133046, K 138136, K 143460, K 143477, 2020-2.2.1-ED-2021-00181, and TKP2021-NKTA-64 (Hungary); the Council of Science and Industrial Research, India; the Latvian Council of Science; the Ministry of Education and Science, project no. 2022/WK/14, and the National Science Center, contracts Opus 2021/41/B/ST2/01369 and 2021/43/B/ST2/01552 (Poland); the Funda\c{c}\~ao para a Ci\^encia e a Tecnologia, grant CEECIND/01334/2018 (Portugal); the National Priorities Research Program by Qatar National Research Fund; MCIN/AEI/10.13039/501100011033, ERDF ``a way of making Europe", and the Programa Estatal de Fomento de la Investigaci{\'o}n Cient{\'i}fica y T{\'e}cnica de Excelencia Mar\'{\i}a de Maeztu, grant MDM-2017-0765 and Programa Severo Ochoa del Principado de Asturias (Spain); the Chulalongkorn Academic into Its 2nd Century Project Advancement Project, and the National Science, Research and Innovation Fund via the Program Management Unit for Human Resources \& Institutional Development, Research and Innovation, grant B05F650021 (Thailand); the Kavli Foundation; the Nvidia Corporation; the SuperMicro Corporation; the Welch Foundation, contract C-1845; and the Weston Havens Foundation (USA).

%% file: atlas_authlist.tex
 
\begin{flushleft}
\hypersetup{urlcolor=black}
{\Large The ATLAS Collaboration}

\bigskip

\AtlasOrcid[0000-0002-6665-4934]{G.~Aad}$^\textrm{\scriptsize 99}$,
\AtlasOrcid[0000-0002-5888-2734]{B.~Abbott}$^\textrm{\scriptsize 116}$,
\AtlasOrcid[0000-0002-7248-3203]{D.C.~Abbott}$^\textrm{\scriptsize 100}$,
\AtlasOrcid[0000-0002-2788-3822]{A.~Abed~Abud}$^\textrm{\scriptsize 34}$,
\AtlasOrcid[0000-0002-1002-1652]{K.~Abeling}$^\textrm{\scriptsize 53}$,
\AtlasOrcid[0000-0002-2987-4006]{D.K.~Abhayasinghe}$^\textrm{\scriptsize 92}$,
\AtlasOrcid[0000-0002-8496-9294]{S.H.~Abidi}$^\textrm{\scriptsize 27}$,
\AtlasOrcid[0000-0002-9987-2292]{A.~Aboulhorma}$^\textrm{\scriptsize 33e}$,
\AtlasOrcid[0000-0001-5329-6640]{H.~Abramowicz}$^\textrm{\scriptsize 148}$,
\AtlasOrcid[0000-0002-1599-2896]{H.~Abreu}$^\textrm{\scriptsize 147}$
% ...
\bigskip
\\

% ...
$^{27}$Physics Department, Brookhaven National Laboratory, Upton NY; United States of America.\\
% ...
$^{33}$$^{(a)}$Facult\'e des Sciences Ain Chock, R\'eseau Universitaire de Physique des Hautes Energies - Universit\'e Hassan II, Casablanca;$^{(b)}$Facult\'{e} des Sciences, Universit\'{e} Ibn-Tofail, K\'{e}nitra;$^{(c)}$Facult\'e des Sciences Semlalia, Universit\'e Cadi Ayyad, LPHEA-Marrakech;$^{(d)}$LPMR, Facult\'e des Sciences, Universit\'e Mohamed Premier, Oujda;$^{(e)}$Facult\'e des sciences, Universit\'e Mohammed V, Rabat; Morocco.\\
$^{34}$CERN, Geneva; Switzerland.\\
% ...
$^{53}$II. Physikalisches Institut, Georg-August-Universit\"{a}t G\"ottingen, G\"ottingen; Germany.\\
% ...
$^{92}$Department of Physics, Royal Holloway University of London, Egham; United Kingdom.\\
% ...
$^{99}$CPPM, Aix-Marseille Universit\'e, CNRS/IN2P3, Marseille; France.\\
$^{100}$Department of Physics, University of Massachusetts, Amherst MA; United States of America.\\
% ...
$^{116}$Homer L. Dodge Department of Physics and Astronomy, University of Oklahoma, Norman OK; United States of America.\\
% ...
$^{147}$Department of Physics, Technion, Israel Institute of Technology, Haifa; Israel.\\
$^{148}$Raymond and Beverly Sackler School of Physics and Astronomy, Tel Aviv University, Tel Aviv; Israel.\\
% ...
$^{a}$ Also Affiliated with an institute covered by a cooperation agreement with CERN.\\
$^{b}$ Also at Borough of Manhattan Community College, City University of New York, New York NY; United States of America.\\
$^{c}$ Also at Bruno Kessler Foundation, Trento; Italy.\\
$^{d}$ Also at Center for High Energy Physics, Peking University; China.\\
$^{e}$ Also at Centro Studi e Ricerche Enrico Fermi; Italy.\\
% ...

\end{flushleft}


%% file: TOP-18-014-public-authorlist.tex
{\Large The CMS Collaboration}
\bigskip
\newskip{\cmsinstskip} \cmsinstskip=0pt plus 4pt

\cmsinstitute{Yerevan Physics Institute, Yerevan, Armenia}
{\tolerance=6000
A.~Tumasyan\cmsAuthorMark{1}\cmsorcid{0009-0000-0684-6742}
\par}
\cmsinstitute{Institut f\"{u}r Hochenergiephysik, Vienna, Austria}
{\tolerance=6000
W.~Adam\cmsorcid{0000-0001-9099-4341}, J.W.~Andrejkovic, T.~Bergauer\cmsorcid{0000-0002-5786-0293}, S.~Chatterjee\cmsorcid{0000-0003-2660-0349}, M.~Dragicevic\cmsorcid{0000-0003-1967-6783}, A.~Escalante~Del~Valle\cmsorcid{0000-0002-9702-6359}, R.~Fr\"{u}hwirth\cmsAuthorMark{2}\cmsorcid{0000-0002-0054-3369}, M.~Jeitler\cmsAuthorMark{2}\cmsorcid{0000-0002-5141-9560}, N.~Krammer\cmsorcid{0000-0002-0548-0985}, L.~Lechner\cmsorcid{0000-0002-3065-1141}, D.~Liko\cmsorcid{0000-0002-3380-473X}, I.~Mikulec\cmsorcid{0000-0003-0385-2746}, P.~Paulitsch, F.M.~Pitters, J.~Schieck\cmsAuthorMark{2}\cmsorcid{0000-0002-1058-8093}, R.~Sch\"{o}fbeck\cmsorcid{0000-0002-2332-8784}, D.~Schwarz\cmsorcid{0000-0002-3821-7331}, S.~Templ\cmsorcid{0000-0003-3137-5692}, W.~Waltenberger\cmsorcid{0000-0002-6215-7228}, C.-E.~Wulz\cmsAuthorMark{2}\cmsorcid{0000-0001-9226-5812}
% ...
\par}
\vskip\cmsinstskip
\dag:~Deceased\\
$^{1}$Also at Yerevan State University, Yerevan, Armenia\\
$^{2}$Also at TU Wien, Vienna, Austria\\
% ...
